\documentclass[11pt]{article}
\usepackage{moriond,amsmath,amssymb,graphicx,epsfig,amsthm,
amsfonts,color}
\usepackage[small,sl]{caption}

\def\beq{\begin{equation}}
\def\eeq{\end{equation}}
\def\bea{\begin{eqnarray}}
\def\eea{\end{eqnarray}}

\begin{document}
\hfill Ref.SISSA 32/2002/EP
\vspace*{3.8cm}
\title{STRUCTURE OF NEUTRINO MASS MATRIX}

\author{ M. FRIGERIO }

\address{INFN, Section of Trieste {\rm and}
International School of Advanced Studies\\ (SISSA/ISAS),
Via Beirut 4, 34014 Trieste, Italy}

\maketitle\abstracts{We reconstruct the neutrino mass matrix 
in the flavor basis, using experimental data on neutrino oscillations. 
Both normal and inverted hierarchy (ordering) of the mass spectrum
are consi\-dered. The dependence of the matrix structure on CP 
violating Majorana phases $\rho$ and $\sigma$ is studied.
The results are shown in $\rho-\sigma$ plots. We find 
the possible matrix structures in the different regions of 
parameters and discuss the perspectives to further restrict 
the matrix structure in future experiments.}


The data on neutrino oscillation experiments \cite{exp} constrain the mixing angles in the leptonic sector and the neutrino mass squared differences. The origin of masses and mixing of neutrinos and the possible underlying flavor symmetries are hidden in the structure of the mass \mbox{matrix \cite{theo}}. We will reconstruct the Majorana mass matrix for three neutrinos in the flavor basis, using the experimental input.  We will stress the importance of CP violating Majorana phases for the matrix structure, both in the case of a mass spectrum with normal hierarchy (ordering) \cite{MA} and in that of inverted hierarchy (ordering) \cite{MA2}.


The neutrino mass matrix in flavor basis, $M$, can be written as
\begin{equation}
M = U^*M^{diag}U^{\dag}~,
\label{matr}
\end{equation}
where
$M^{diag} \equiv Diag(m_1e^{-2i\rho},~m_2,~m_3e^{-2i\sigma})$.
Here $m_i$ are the moduli of the neutrino mass eigenvalues;
$\rho$ and $\sigma$ are the two CP violating Majorana phases,
varying between $0$ and $\pi$.
The standard para\-meterization for the mixing matrix $U$ is:
\begin{equation}
U =
\left( \begin{array}{ccc}
 c_{13}c_{12} & s_{12}c_{13} & s_{13} e^{-i\delta}
\\ -s_{12}c_{23}-s_{23}s_{13}c_{12} e^{i\delta} &
    c_{23}c_{12}-s_{23}s_{13}s_{12} e^{i\delta}
& s_{23}c_{13} \\ s_{23}s_{12}-s_{13}c_{23}c_{12} e^{i\delta}
 & -s_{23}c_{12}-s_{13}s_{12}c_{23} e^{i\delta} & c_{23}c_{13}
\end{array} \right)\;,
\label{U}
\end{equation}
where $c_{ij} \equiv \cos \theta_{ij}$, $s_{ij} \equiv \sin \theta_{ij}$
and $\delta$ is the CP violating Dirac phase. The mixing angles
vary between $0$ and $\pi/2$ and $\delta$ between $0$ and $2\pi$.

We will consider the absolute values of the six independent matrix elements, which are functions of nine parameters:
$m_{\alpha\beta}=|M_{\alpha\beta}(m_i,\theta_{ij},\delta,\rho,\sigma)|$, 
$\alpha,\beta=e,\mu,\tau$.
Experimental data on neutrino oscillations \cite{exp} give five constraints on the parameter space ($90\%$ C.L.):
\beq
\begin{array}{l}
m_2^2-m_1^2 \equiv \Delta m^2_{sol}=
(5^{+15}_{-3}) \cdot 10^{-5} {\rm eV}^2\;;\qquad
|m_3^2-m_2^2|
\equiv \Delta m^2_{atm}=
(2.5^{+1.5}_{-0.9}) \cdot 10^{-3} {\rm eV}^2\;;\\
\tan^2 \theta_{12}=  0.35^{+0.3}_{-0.1}\;;\qquad
\tan \theta_{23} = 1 ^{+ 0.4}_{-0.3}\;;\qquad
s_{13} \lesssim 0.2\;
\end{array}
\label{data}
\end{equation}
(we use the LMA MSW solution of the solar neutrino problem). 
The absolute mass scale and the three CP violating phases 
are not constrained by oscillation data.


In order to study the dominant structure of the mass matrix, 
we will neglect ${\cal O}(s_{13})$ terms with respect to 
${\cal O}(1)$ terms. 
Notice, however, that some matrix elements can be of order $s_{13}$.
Therefore, the subdominant structure of the matrix, formed by small elements,
cannot be studied in this approximation. A more detailed analysis 
can be found in a previous work \cite{MA}.
Using Eqs.(\ref{matr}) and (\ref{U}) and defining $k\equiv m_1/m_2,\;r\equiv m_3/m_2$, for $s_{13}=0$ we get:
\beq
\frac{m}{m_2}=\left(
\begin{array}{ccc}
z & c_{23} y & s_{23} y\\
\dots & |c_{23}^2  x + s_{23}^2 r e^{-2i\sigma_x}| &
s_{23} c_{23} |- x + r e^{-2i\sigma_x}|\\
\dots & \dots & |s_{23}^2 x + c_{23}^2 r e^{-2i\sigma_x}|
\end{array}
\right) \;,
\label{general}
\end{equation}
where $z\equiv |s_{12}^2+c_{12}^2 k e^{-2i\rho}|$, 
$y\equiv c_{12}s_{12}|1-ke^{-2i\rho}|$, 
$X\equiv c_{12}^2+s_{12}^2 k e^{-2i\rho}$, 
$x\equiv |X|$, $\sigma_x \equiv \sigma + {\arg X} /2$.
The solar mass difference can be neglected if $k$ is approximately 
equal to $1$ ($k>0.95$  
for $m_1^2 \gtrsim 9\Delta m^2_{sol}$). 
In the limit $k=1$ ($m_1=m_2$), 
the mass matrix (\ref{general}) becomes:  
\beq
\frac{m}{m_2}=\left(
\begin{array}{ccc}
x & c_{23} \sqrt{1-x^2} & s_{23} \sqrt{1-x^2}\\
\dots & |c_{23}^2  x + s_{23}^2 r e^{-2i\sigma_x}| &
s_{23} c_{23} |- x + r e^{-2i\sigma_x}|\\
\dots & \dots & |s_{23}^2 x + c_{23}^2 r e^{-2i\sigma_x}|
\end{array}
\right) \;.
\label{matrix}
\end{equation}
Different mass spectra 
are described by Eq.(\ref{matrix}), depending on the value of $r$: 
inverted hierarchy (ordering) for $0\le r<1$, 
degeneracy for $r=1$, normal ordering for $1<r\lesssim 2\div 3$. 
For larger values of $r$ (normal hierarchy), the approximation $k=1$ 
is no longer valid 
and one should use Eq.(\ref{general}).

As regards the other parameters in Eq.(\ref{matrix}), 
using Eq.(\ref{data}) we get $c_{23}^2=0.5\pm 0.15$, 
$s_{23}^2=0.5\mp 0.15$, $x\in [\cos2\theta_{12},1]=[0.2\div 0.6,1]$. 
The value of $x$ in this interval is fixed by $\rho$. 
The phase $\sigma_x$ is unconstrained, because $\sigma$ is.
Therefore, the matrix structure can depend significantly 
on the values of the two Majorana phases. 
CP is conserved if $\rho$ and $\sigma$ are equal to
$0$ or $\pi/2$ (definite CP parities 
of the three neutrino eigenvalues).
To our knowledge, there is not a theoretical principle which 
privileges CP conserving values of the Majorana phases $\rho$ 
and $\sigma$ over CP violating ones. Therefore, it is important 
to consider both the structures which correspond 
to definite CP parity assignments and the others. In particular, 
certain ``symmetric'' features  of the mass matrix imply CP violation.
For example, the elements $m_{\mu\mu}$, $m_{\mu\tau}$ 
and $m_{\tau\tau}$, in Eq.(\ref{matrix}), 
are equal only for  $\theta_{23}=\pi/4$ and 
$\sigma_x = \pi/4, 3\pi/4$. 
The elements $m_{ee}$, $m_{e\mu}$ and $m_{e\tau}$ 
are equal for $\theta_{23}=\pi/4$ and $x=1/\sqrt{3}$.

\begin{figure}[p]
\begin{center}
\epsfig{figure=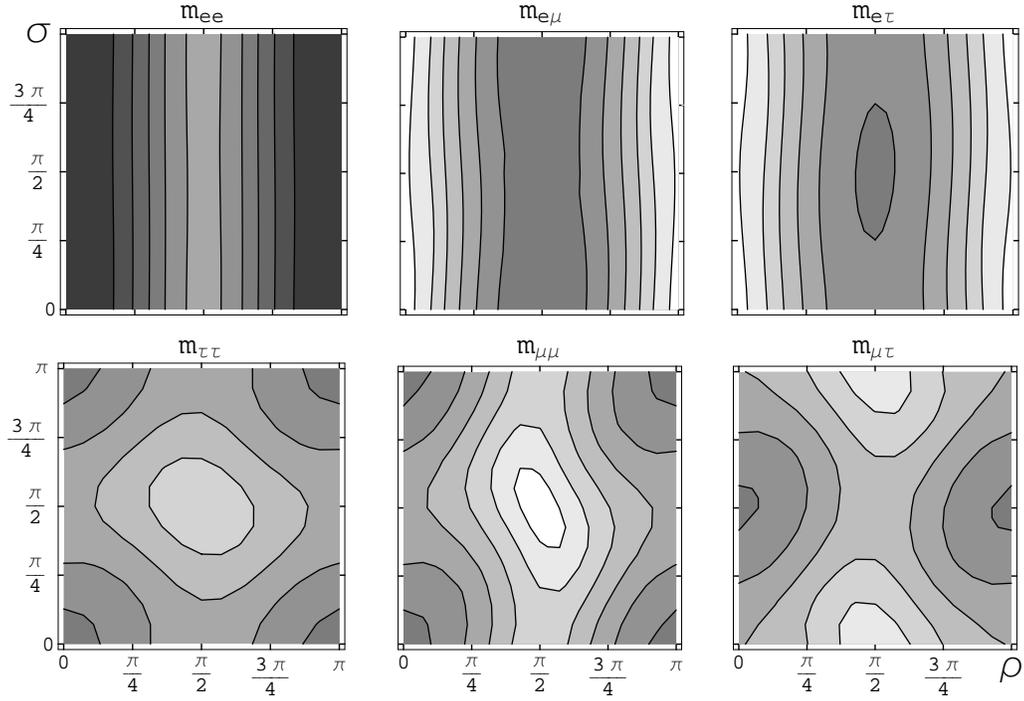,height=3.7in, 
bbllx=70, bblly=375, bburx=560, bbury=710}
\caption{The $\rho-\sigma$ plots for inverted hierarchical spectrum,
with $m_3=0.01$ eV, corresponding to $r\approx 0.2$. The contours 
are shown of constant mass $m=(0.1,\,0.2,\,\dots,\,0.9)m^{max}$, 
where $m^{max}=0.05$ eV, so that the white regions correspond 
to the mass interval ($0-0.005$) eV and the darkest ones 
to ($0.045-0.050$) eV. 
We take $\Delta m^2_{sol}=5 \cdot 10^{-5} {\rm eV}^2$, 
$\Delta m^2_{atm}= 2.5 \cdot 10^{-3} {\rm eV}^2$ and 
$\tan^2\theta_{12}=0.35$, $\tan\theta_{23}=1$, $s_{13}=0.1$, $\delta=0$.
\label{fig1}}
\end{center}
\end{figure}

\begin{figure}[p]
\begin{center}
\epsfig{figure=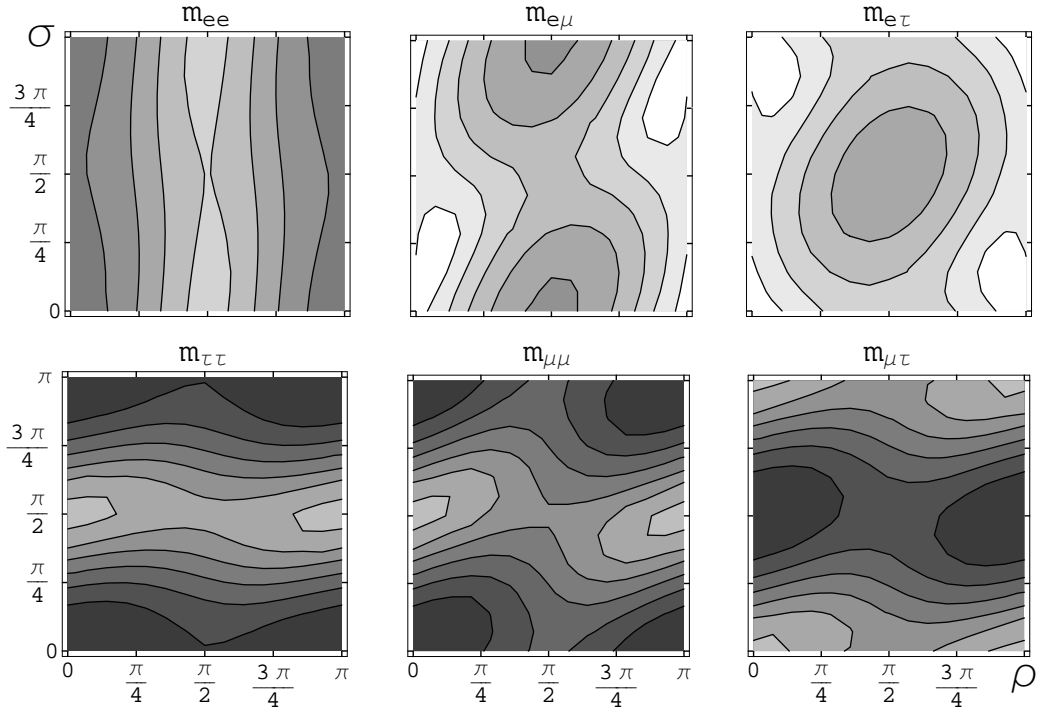,height=3.7in, bbllx=70, bblly=460, bburx=515, bbury=760}
\caption{The same as in Fig.\ref{fig1}, 
but for normal hierarchical spectrum,
with $m_1=0.025$ eV, corresponding to $r\approx 2$.
In this case $m^{max}=0.04$ eV, so that the white regions correspond 
to the mass interval ($0-0.004$) eV and the darkest ones 
to ($0.036-0.040$) eV.
\label{fig2}}
\end{center}
\end{figure}

In Figs.\ref{fig1} and \ref{fig2}, we show contours of constant values 
of $m_{\alpha\beta}$ in the plane of the Majorana phases, in the 
case $r\approx 0.2$ and $r\approx 2$ respectively. Similar plots 
for other values of $r$ have been given in a previous work \cite{MA}. 
We take a value of $s_{13}$ different from zero, 
so that one can see the effect of ${\cal O}(s_{13})$
corrections, neglected in Eqs.(\ref{general}) and (\ref{matrix}).
All features of the mass matrix can be easily extracted from these plots. 
Three kinds of structures can be identified:
\begin{itemize}
\item
If some elements are very small (white regions in the 
$\rho-\sigma$ plots) and others are of order one (dark regions), 
we can say that the mass matrix has a ``{\it hierarchical structure}''.
Notice that, taking into account the range of variation of 
$r,x,y,z,\sigma_x$ and $\theta_{23}$ in Eqs.(\ref{general}) and 
(\ref{matrix}), 
each mass matrix element can be zero. Therefore many hierarchical 
structures are possible.
\item
On the other hand, all the elements can be equal (for 
$\theta_{23}=\pi/4$, $\sigma_x=\pi/4$, $r=1$, $x=1/\sqrt{3}$). 
In this case the mass matrix has a ``{\it democratic structure}''
(remember that the complex phases of the matrix elements can be different).
In the $\rho-\sigma$ plots, this 
corresponds to the same gray level for all the elements.
\item Between the extreme situations of hierarchical and 
democratic structures, there are many possible 
mass matrices with some specific ordering in the values of matrix 
elements: we name it ``{\it ordering structure}''.
\end{itemize}

One may ask what the perspectives are to select a unique structure 
for the neutrino mass matrix. In order to do that, 
one has to fix the mass spectrum and the Majorana phases.
There is a concrete chance of measuring the absolute mass scale 
and the phase $\rho$; this can be done improving the 
kinematic bound on $m_{\nu_e}$ 
($m_{\nu_e}\equiv (M^\dag M)^{1/2}<2.2$ eV ($95\%$ C.L.) \cite{beta})
and using neutrinoless $2\beta$-decay experiments. 
The probability of this decay is proportional to 
$m_{ee}^2 \approx (m_2 z)^2$.
The updated upper bound \cite{2beta} is $m_{ee}<0.35$ eV ($90\%$ C.L.).
If $m_2>0.35$ eV (degenerate spectrum, $r\approx 1$), this bound 
is sensitive to the value of $\rho$.
The $\rho-\sigma$ plots and Eq.(\ref{matrix}) show that 
a constraint on $m_{ee}$ 
restricts substantially also the other matrix elements. 
The measure of $m_{ee}$ would determine the 
other elements if their dependence on $\sigma$ is weak, 
that is for $r\ll 1$ (inverted hierarchy) and $r\gg 1$ (normal hierarchy). 
If $r\sim 1$, instead, the elements $m_{\mu\mu}$, $m_{\mu\tau}$ 
and $m_{\tau\tau}$ are strongly $\sigma$-dependent. 
The determination of $\sigma$ is out of the reach of 
the forthcoming experiments.

In conclusion, we have shown that the neutrino mass matrix strongly depends
on the CP violating Majorana phases. We have found that, depending on 
the value of $\rho$ and $\sigma$, the matrix can have hierarchical, 
democratic or ordering structures.
Neutrinoless $2\beta$-decay experiments are sensitive to the phase $\rho$.
We have shown how the determination of $\rho$ would strongly 
constrain the matrix structure.


\section*{Acknowledgments}
I would like to thank the organizers of the XXXVIIth Rencontres de Moriond 
for financial support. Special thanks to Alexei Smirnov 
for the useful suggestions during the preparation of these Proceedings.

\end{document}